\pdfoutput=1
\documentclass[sigconf,natbib=true,anonymous=false]{acmart}

\setcopyright{none}
\renewcommand\footnotetextcopyrightpermission[1]{}
\usepackage{multirow, makecell}
\usepackage{xcolor}
\usepackage{colortbl}
\usepackage{dsfont}

\usepackage[T1]{fontenc}

\usepackage[hang,flushmargin]{footmisc}

\usepackage{url}            
\usepackage{booktabs}       
\usepackage{amsfonts}       
\usepackage{nicefrac}       
\usepackage{microtype}      
\usepackage{amsmath}
\usepackage{enumitem}
\usepackage{graphicx}       
\usepackage{caption}
\usepackage{subcaption}
\usepackage{threeparttable}
\usepackage{color, colortbl}
\usepackage{multirow}
\usepackage{makecell}
\usepackage{xspace}
\usepackage{dsfont}
\usepackage{listings}
\usepackage{balance}
\usepackage{gensymb}
\usepackage{hyperref}

\usepackage[frozencache=true,cachedir=minted-cache]{minted} 
\definecolor{LightGray}{gray}{0.9}
\newcommand{\umb}{UM\-BRELA\xspace}

\PassOptionsToPackage{dvipsnames}{xcolor}

\definecolor{magicmint}{rgb}{0.67, 0.94, 0.82}
\definecolor{Box3Color}{RGB}{255, 214, 193}
\PassOptionsToPackage{dvipsnames}{xcolor}
\usepackage{ragged2e}

\lstdefinestyle{json}{
    basicstyle=\footnotesize\ttfamily,
    numberstyle=\tiny,
    showstringspaces=false,
    breaklines=true,
    breakindent=0em,
    frame=bt,
    numbers=none,
    xleftmargin=.03in,
    xrightmargin=.03in,
    backgroundcolor=\color{Box3Color},
}

\newcommand\ignore[1]{}

\begin{document}

\title{UMBRELA: UMbrela is the (Open-Source Reproduction of the) Bing RELevance Assessor}

\author{Shivani Upadhyay}
\email{sjupadhyay@uwaterloo.ca}
\affiliation{%
  \institution{University of Waterloo}
  \city{Waterloo}
  \country{Canada}
  }

\author{Ronak Pradeep}
\email{rpradeep@uwaterloo.ca}
\affiliation{%
  \institution{University of Waterloo}
  \city{Waterloo}
  \country{Canada}
}

\author{Nandan Thakur}
\email{nandan.thakur@uwaterloo.ca}
\affiliation{%
  \institution{University of Waterloo}
  \city{Waterloo}
  \country{Canada}
}

\author{Nick Craswell}
\email{nickcr@microsoft.com}
\affiliation{%
  \institution{Microsoft}
  \city{Seattle}
  \country{USA}
}

\author{Jimmy Lin}
\email{jimmylin@uwaterloo.ca}
\affiliation{%
  \institution{University of Waterloo}
  \city{Waterloo}
  \country{Canada}
}

\renewcommand{\shortauthors}{}
\pagestyle{empty}

\begin{abstract}
Copious amounts of relevance judgments are necessary for the effective training and accurate evaluation of retrieval systems.
Conventionally, these judgments are made by human assessors, rendering this process expensive and laborious.
A recent study by Thomas et al.\ from Microsoft Bing suggested that large language models (LLMs) can accurately perform the relevance assessment task and provide human-quality judgments, but unfortunately their study did not yield any reusable software artifacts.
Our work presents \umb (a recursive acronym that stands for \textbf{UM}brela is the \textbf{B}ing \textbf{REL}evance \textbf{A}ssessor), an open-source toolkit that reproduces the results of Thomas et al.\ using OpenAI's GPT-4o model and adds more nuance to the original paper.
Across Deep Learning Tracks from TREC 2019 to 2023, we find that LLM-derived relevance judgments correlate highly with rankings generated by effective multi-stage retrieval systems.
Our toolkit is designed to be easily extensible and can be integrated into existing multi-stage retrieval and evaluation pipelines, offering researchers a valuable resource for studying retrieval evaluation methodologies.
\umb will be used in the TREC 2024 RAG Track to aid in relevance assessments, and we envision our toolkit becoming a foundation for further innovation in the field.
\umb is available at \url{https://github.com/castorini/umbrela}.
\end{abstract}

\maketitle

\section{Introduction}

Accurate relevance labels are crucial for training and evaluating retrieval systems.
Typically, these labels are assigned by trained human assessors, for example, retired intelligence analysts in the case of TREC or editors hired by a web search engine company.
Given a search query, a set of results, and a description of the information need, these humans assessors determine the relevance of those results.
However, human assessments can be time-consuming, labor-intensive, and costly.
Even after allocating these resources, assessments often remain inaccurate and misaligned with the searcher's intentions, particularly in the case where the information need did not arise from the assessor.
They may fail to grasp the intent behind a search, resulting in flawed relevance judgments.

The emergence of large language models (LLMs) such as GPT-4~\cite{gpt4} and Gemini~\cite{gemini} has presented a distinctive opportunity to automate the intricate task of data annotation~\cite{alizadeh23opensource,chiang23llm,gilardi23chatgpt,liu23geval,tornberg23chatgpt,törnberg2024best,tan2024large,annollm}.
\citet{bing} showcased capabilities of LLMs to understand searcher intent and to assign relevant labels as ``accurately'' as human assessors.
These models have the potential to serve as an efficient and cost-effective alternative for manual relevance assessment.
Moreover, LLMs may actually ``outperform'' assessors with limited knowledge, who might not fully grasp the search intent, thereby leading to imprecise relevance judgments. 
In addition, relevance judgments facilitated by LLMs can be employed to ``fill holes'' left by incomplete judgments, thereby contributing to a more thorough and accurate evaluation of retrieval systems~\cite{plugholes, oneshot}.

In this paper, we reproduced the work of~\citet{bing} and repackage our efforts into \umb, an open-source toolkit for using LLMs as relevance assessors.
UMbrela is a recursive acronym that stands for \textbf{UM}brela is the \textbf{B}ing \textbf{REL}evance \textbf{A}ssessor.
In our study, we carefully follow the zero-shot DNA (Descriptive, Narrative, and Aspects) prompting technique described by~\citet{zero-shot} to reproduce the original setup as faithfully as possible.

Our experiments with the TREC Deep Learning (DL) Tracks from 2019--2023~\cite{trec19, trec20, trec21, trec2022, trec2023} verified the claims by~\citet{bing} about LLM assessment being an effective alternative to human-based assessment, using OpenAI's latest model GPT-4o.
We have expanded our validation study to include correlation statistics that compare the effectiveness of retrieval systems based on human and LLM judgments.
High correlations affirm the effectiveness of LLMs as relevance assessors.
As highlighted by~\citet{bing}, LLMs possess the ability to understand user needs and assign relevance labels with veracity comparable to human assessors.

At a high level, \umb takes a query and a set of passages as input, and based on the LLM configuration, attempts to ``understand'' the search intent and labels the passages with different relevance grades.
Our main contributions are as follows:

\begin{itemize}[leftmargin=*]

\item We reproduced results from~\citet{bing}, which lacked reusable software components, and created \umb, an open-source toolkit that we share with the broader community.

\item Our tool will be used in the TREC 2024 RAG Track\footnote{\url{https://cs.uwaterloo.ca/~jimmylin/publications/Lin_etal_TREC2023-planning.pdf}} to showcase its practical utility for automatic relevance assessment.
Through this deployment, we hope to contribute to advances in retrieval evaluation methodologies.

\end{itemize}

\section{Related Work}

The substantial effort required for manual test data preparation in large-scale testing of retrieval systems has consistently encouraged researchers to develop automated techniques~\cite{dietz2022wikimarks,dietz2020humans}. 
In previous work, automated IR evaluations using LLMs have combined various prompting techniques such as zero-shot, one-shot, or few-shot learning.
By providing detailed instruction, defining roles and multiple judges, researchers have showcased a continuous increase in the alignment between human and the predicted assessments~\cite{faggioli23perspectives, oneshot, bing, plugholes}. 
LLMs, when provided with proper guidance, can effectively ``understand'' the relation between a given query and passage and accurately determine its relevance.

In the domain of text annotations in general, both proprietary and open-source LLMs have shown remarkable effectiveness~\cite{zhu2023can,savelka2023can,annollm,alizadeh23opensource,alizadeh2024opensource}. 
The enhanced capabilities of LLMs and various prompting techniques have presented a strong substitute for manual text annotations.

\section{Methodology}

The main idea behind \umb is to leverage LLMs for relevance assessment, reproducing the work of~\citet{bing}.

\begin{table}[t]
  \begin{tabular}{l|>{\centering}p{0.9cm}>{\centering}p{0.8cm}|>{\centering}p{0.8cm}>{\centering}p{0.8cm}>{\centering}p{0.8cm}>{\centering\arraybackslash}p{0.8cm}}
    \toprule
    \centering \textbf{qrels} & \multicolumn{6}{c}{\textbf{Total counts for}}\\
    \centering \textbf{track} &\textbf{runs}&\textbf{topics}&\textbf{label 0}&\textbf{label 1}&\textbf{label 2}&\textbf{label 3}\\
    \midrule
    DL 2019 & 36 & 43 & 5158 & 1601 & 1804 & 697\\
    DL 2020 & 59 & 54 & 7780 & 1940 & 1020 & 646\\
    DL 2021 & 62 & 53 & 4338 & 3063 & 2341 & 1086\\
    DL 2022 & 60 & 76 & 286459 & 52218 & 46080 & 1659\\
    $\text{DL 2022}^*$ & 60 & 76 & 12892 & 6192 & 3053 & 1385\\
    DL 2023 & 35 & 82 & 13866 & 4372 & 2259 & 1830\\
    $\text{DL 2023}^*$ & 35 & 82 & 11618 & 3774 & 1942 & 1544\\
  \bottomrule
  \end{tabular}
  \vspace{2mm}
  \caption{Summary statistics for the Deep Learning Tracks from TREC 2019--2023. The total count of topics represents only those that have qrels; $*$ represents counts after applying the pre-processing step to remove near-duplicate passages, as explained in Section~\ref{sec:deduped}.}
  \label{tab:dl-tracks}
 \vspace{-0.25cm}
\end{table}

\subsection{DL TREC Judgments}

To demonstrate LLM capabilities for generating relevance assessment, we took human relevance judgments (also known as qrels) from the Deep Learning Tracks in TREC 2019--2023~\cite{trec19, trec20, trec21, trec2022, trec2023} and re-assessed them with \umb. 
These judgments were originally provided by human NIST assessors who were given the query and the following relevance scale~\cite{trec19}:

\begin{itemize}
\item [0] Irrelevant: The passage has nothing to do with the query
\item [1] Related: The passage seems related to the query but does not answer it
\item [2] Highly relevant: The passage has some answer for the query, but the answer may be a bit unclear, or hidden amongst extraneous information
\item [3] Perfectly relevant: The passage is dedicated to the query and contains the exact answer
\end{itemize}

\noindent Based on their understanding and general knowledge, the human assessors assigned relevance labels to passages.
Thus, we consider these as ``gold'' labels.
See Table~\ref{tab:dl-tracks} for summary statistics.

\paragraph{Near-duplicate handling for DL 2022-23}
\label{sec:deduped}

The qrels for the TREC 2022 and 2023 Deep Learning Tracks include relevance label propagation for near-duplicate passages that exist within the corpus.
Thus, the number of relevance judgments that were actually assigned by humans is much smaller than the statistics in Table~\ref{tab:dl-tracks} suggest (we provide figures both with and without de-duplication).
For the purposes of this study, we have excluded these near-duplicate passages from both the judgments and the submission results prior to conducting our calculations.
To accomplish this, we consulted the list of near-duplicate passages compiled by NIST, which provides the near-duplicates in the form of clusters and a ``canonical'' passage for each cluster.
Specifically, we retained only the ``canonical'' passages; all others were eliminated both from the qrels and the retrieved runs.

\subsection{Prompting Details}

Following the prompting techniques described by~\citet{bing}, we also utilize the DNA prompting technique.
Figure~\ref{fig:prompt} shows the exact prompt used in \umb for relevance assessment.

\begin{figure}[t]
\tiny
\justifying
\begin{minted}[fontsize=\footnotesize, frame=lines, frame=single,linenos=false,breaklines,breaksymbol=,escapeinside=||,bgcolor=LightGray]{text}
Given a query and a passage, you must provide a score on an integer scale of 0 to 3 with the following meanings:
0 = represent that the passage has nothing to do with the query, 
1 = represents that the passage seems related to the query but does not answer it, 
2 = represents that the passage has some answer for the query, but the answer may be a bit unclear, or hidden amongst extraneous information and 
3 = represents that the passage is dedicated to the query and contains the exact answer.

Important Instruction: Assign category 1 if the passage is somewhat related to the topic but not completely, category 2 if passage presents something very important related to the entire topic but also has some extra information and category 3 if the passage only and entirely refers to the topic. If none of the above satisfies give it category 0.

Query: {query}
Passage: {passage}

Split this problem into steps:
Consider the underlying intent of the search.
Measure how well the content matches a likely intent of the query (M).
Measure how trustworthy the passage is (T).
Consider the aspects above and the relative importance of each, and decide on a final score (O). Final score must be an integer value only.
Do not provide any code in result. Provide each score in the format of: ##final score: score without providing any reasoning.
\end{minted}
\vspace{-0.5cm}
\caption{Prompt used for relevance assessment.}
\label{fig:prompt}
\vspace{-0.3cm}
\end{figure}
\begin{figure*}[bth]
    \vspace{-0.8cm}
        \centering
        \includegraphics[width=0.33\textwidth]{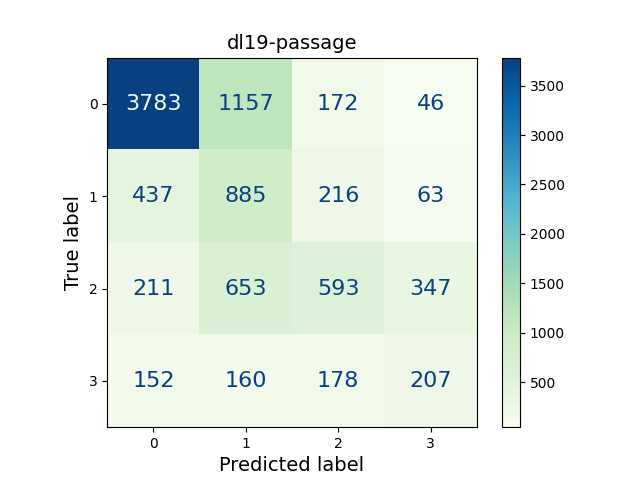}
        \includegraphics[width=0.33\textwidth]{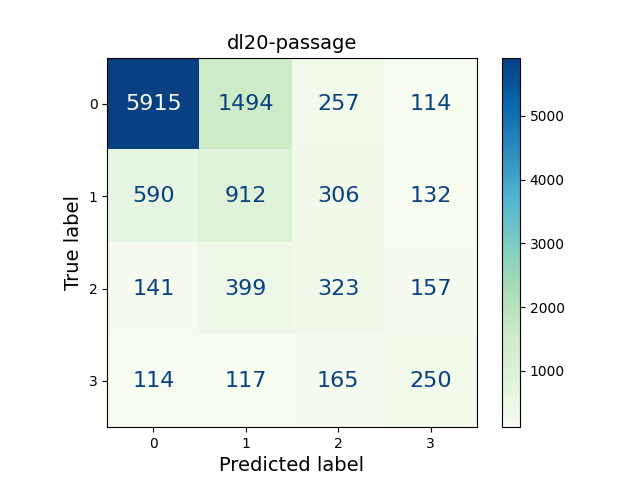} \\
        \vspace{-0.1cm}
        \includegraphics[width=0.33\textwidth]{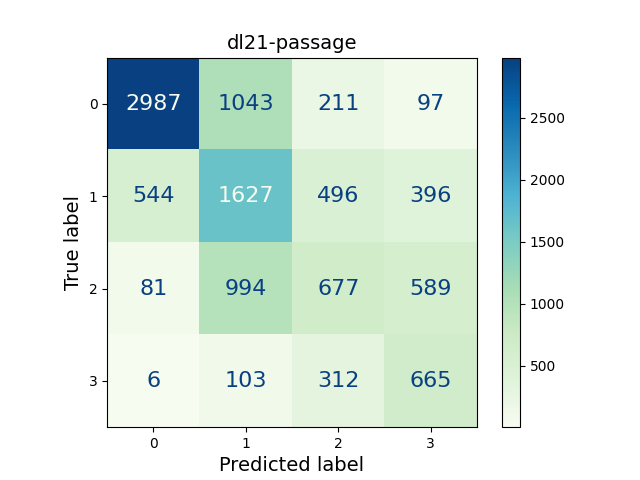}
        \includegraphics[width=0.33\textwidth]{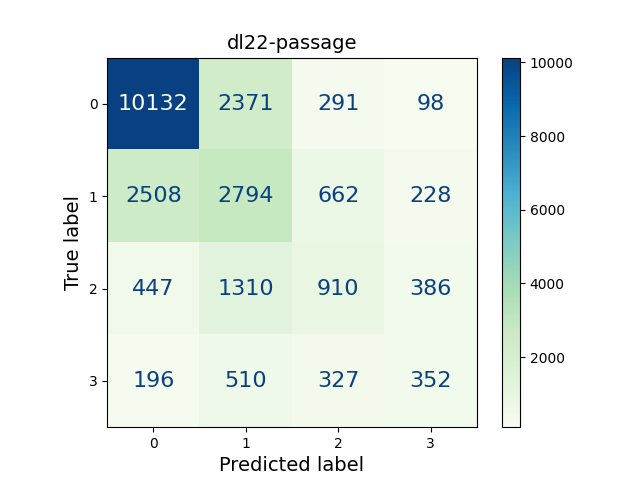}
        \includegraphics[width=0.33\textwidth]{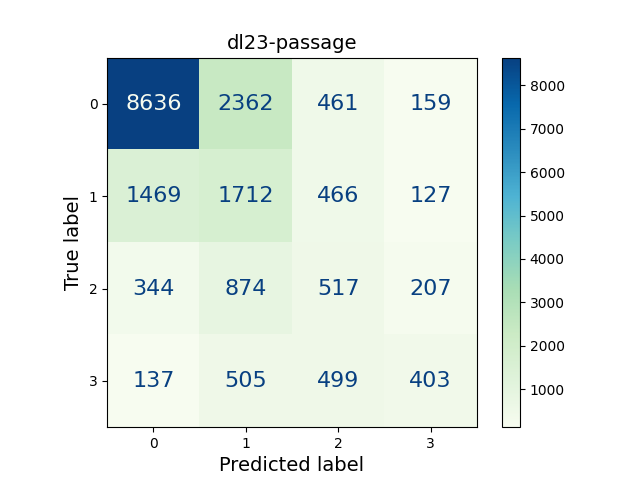}
    \caption{Confusion matrices comparing the original human labels with those generated by the LLM.}
    \label{fig:conf_matrix}
\end{figure*}

The DNA prompt consists of three essential sections:\ descriptive, narrative, and aspects. 
The descriptive and narrative sections help the LLM understand the user query and the passage that it is supposed to label, whereas the aspects section details the step-by-step procedure for guiding the LLM's ``thinking process''.
This procedure essentially breaks down the relevance labeling task into smaller ones and instructs the LLM to develop a better interpretation of it.
The prompt additionally informs the LLM of the expected result structure to be followed.

Following the original paper, we utilize a zero-shot prompting technique for performing our assessments using GPT-4o.
Providing precise instructions assists the LLM in understanding the semantic relation between the query and the passage and thus assigning accurate labels.

\section{Results}

We performed experiments with OpenAI's latest GPT-4o model (via Microsoft Azure) following the parameter configurations provide in~\citet{bing}.
Temperature is set to 0, $\mathrm{top}~p=1$, without using any stopwords, frequency and presence penalty set as 0.5 and 0, respectively.

\begin{table}[t]
  \vspace{-0.1cm}
  \begin{tabular}{l|c c|c c}
    \toprule
    \multirow{2}{*}{\textbf{qrels track}} & \multicolumn{2}{c}
    {\textbf{Cohen $\kappa$}} & \multicolumn{2}{c}
    {\textbf{Correlation}}\\
    &\textbf{four-scale}&\textbf{binary}
    &\textbf{Kendall}&\textbf{Spearman}\\
    &&&\textbf{$(\tau)$}&\textbf{$(\rho)$}\\
    \midrule
    DL 2019 & 0.3613 & 0.4989 & 0.8926 & 0.9736\\
    DL 2020 & 0.3506 & 0.4496 & 0.9435 & 0.9923\\
    DL 2021 & 0.3730 & 0.4917 & 0.9343 & 0.9915\\
    $\text{DL 2022}^*$ & 0.3362 & 0.4217 & 0.8728 & 0.9729\\
    $\text{DL 2023}^*$ & 0.3081 & 0.4176 & 0.9107 & 0.9857\\
  \bottomrule
  \end{tabular}
  \vspace{2mm}
  \caption{Cohen $\kappa$ scores (left) and Kendall and Spearman's correlations between TREC submissions evaluated with ground-truth judgments and \umb (right). We use nDCG@10 as the evaluation metric. Here, $*$ represents the use of a pre-processing step to remove near-duplicate passages as explained in Section~\ref{sec:deduped}.}
  \label{tab:results}
  \vspace{-0.25cm}
\end{table}
\begin{figure*}[tbh]
        \centering
        \includegraphics[width=0.33\textwidth]{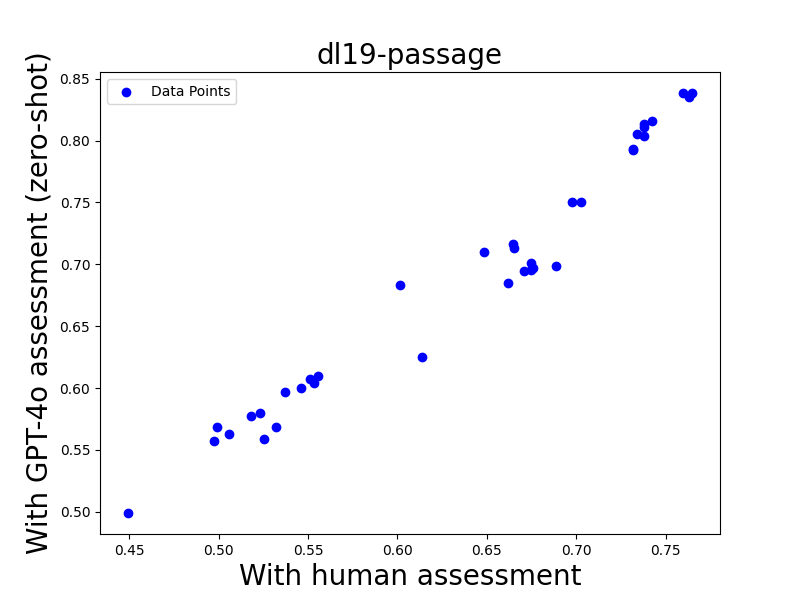}
        \includegraphics[width=0.33\textwidth]{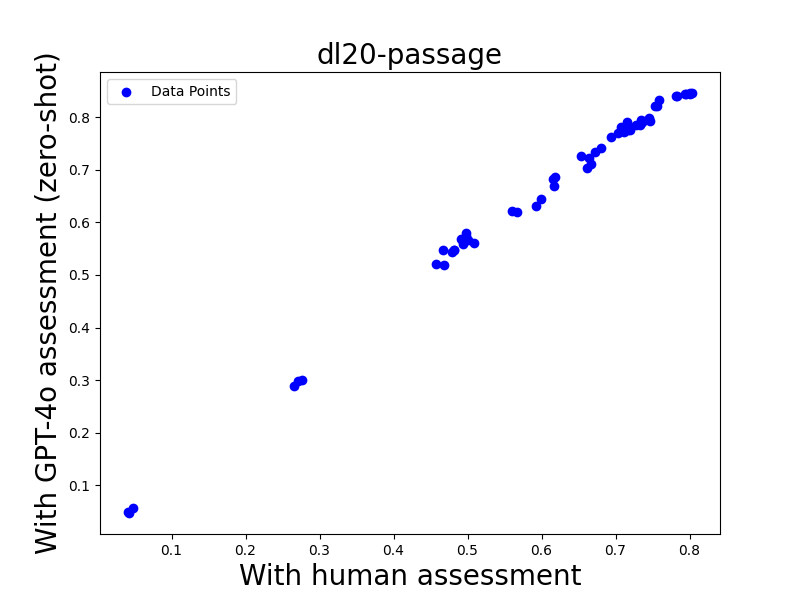} \\
        \vspace{-0.1cm}
        \includegraphics[width=0.33\textwidth]{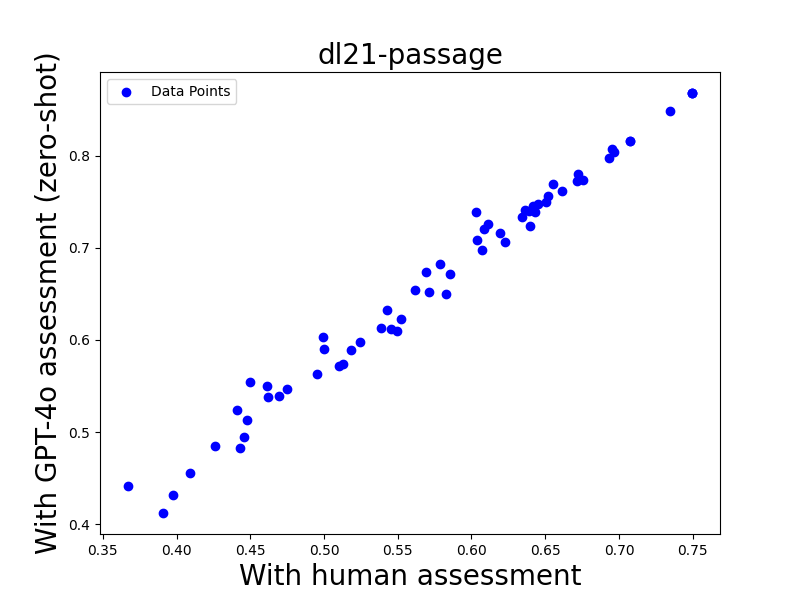}
        \includegraphics[width=0.33\textwidth]{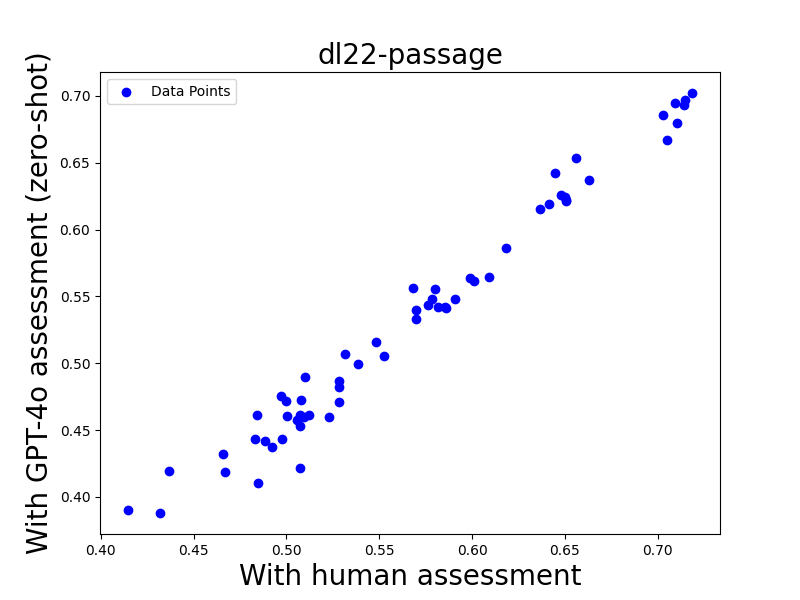}
        \includegraphics[width=0.33\textwidth]{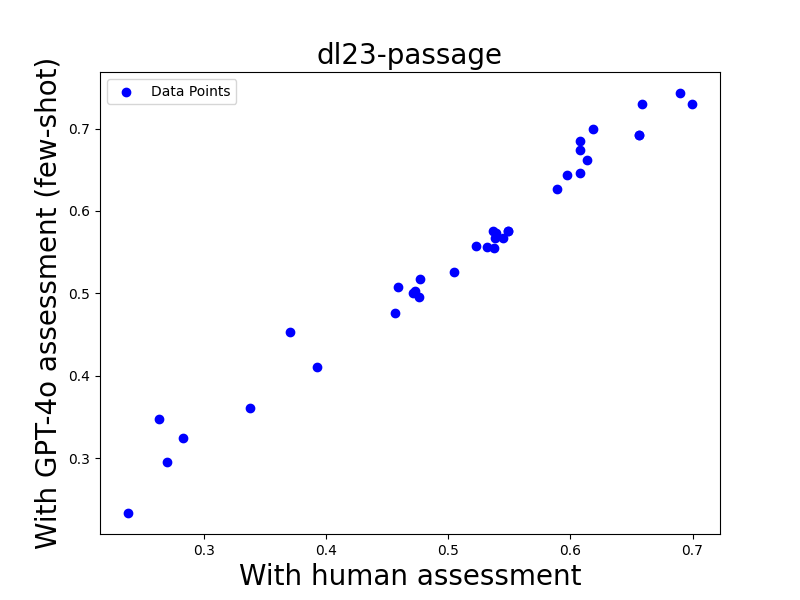}
    \vspace{-0.3cm}

    \caption{Scatter plots for comparing evaluations performed using original human assessments and LLM assessments.}
    \label{fig:scatter_plots}
\end{figure*}

\paragraph{Comparing LLM-based assessments.}
Table~\ref{tab:results} (left) presents Cohen $\kappa$ scores to capture agreement between human and LLM assessments. 
These agreement scores are demonstrated in two variants, the four-scale form which considers all four relevance labels and the binarized form following~\citet{faggioli23perspectives}, where non-relevant (0) and related (1) are merged under ``non-relevant'' and highly relevant (2) and perfectly relevant (3) are merged under the ``relevant'' label.
These correlation values demonstrate fair agreement between human assessors and LLMs judgments.

Figure~\ref{fig:conf_matrix} further compares human-provided ground-truth labels with LLM-based labels using label-wise confusion matrices for all five tracks. 
As expected, LLMs are able to assign appropriate labels to the majority of the judgments across all tracks.
Across these datasets, the LLM was able to predict non-relevant labels with roughly 75\% accuracy, whereas for labels ``related'', ``highly relevant'', and ``perfectly relevant'', the accuracy drops to roughly 50\%, 30\%, and 45\%, respectively (with respect to the human labels).

\paragraph{System evaluation using LLM-based assessments.}
Table~\ref{tab:results} (right) shows the results of Kendall $\tau$ and Spearman $\rho$ correlations between evaluations performed with human ground-truth judgments and judgments provided by the LLM.
In all cases, we compared nDCG@10.
The high correlations among these judgment evaluations show alignment in the retrieval systems' rankings.
Following~\citet{faggioli23perspectives}, Figure~\ref{fig:scatter_plots} demonstrates the effectiveness of TREC DL runs with scatter plots comparing evaluations made using human and LLM judgments.

\paragraph{Case Study:\ TREC DL 2019.}
We further analyzed the discrepancies between LLM judgments and human judgments for TREC DL 2019.
As shown in Figure~\ref{fig:conf_matrix}(a), we observe a high number of misclassifications in the judgments, predicted as label 0 but where their original (human) label is 3.
Based on manual spot-checking, we encountered the following interesting case:

\begin{itemize}[label={},leftmargin=0.25cm]
    \item \textbf{Query:} \textit{what is the daily life of thai people}
    \item \textbf{Passage:} \textit{Thai Flag Meaning: The red stripes represent the blood spilt to maintain Thailand's independence. The white stands for purity and is the color of Buddhism which is the country's main religion.roportions: 2:3. Thai Flag Description: The flag of Thailand consists of five horizontal stripes. The top and bottom are equal-sized red stripes, the middle stripe is blue which is lined above and below by equal-sized white stripes. The blue stripe is double the size of the other four.}
\end{itemize}

\noindent The passage mainly talks about the Thai flag but the query asks about the ``regular life of Thai people''.
Here the topic seems to be somewhat vague and the pair fails to justify (based on our opinion) the highly relevant label assigned by the human assessors.
A few other judgments for this query exhibit similar ambiguity and thus we feel the label 3 to be unwarranted.
In this case, the LLM judgment appears to be more accurate.

Consider another case:

\begin{itemize}[label={},leftmargin=0.25cm]
    \item \textbf{Query:} \textit{medicare's definition of mechanical ventilation}
    \item \textbf{Passage:} \textit{Continuous Positive Airway Pressure (CPAP) Continuous positive airway pressure -- also called CPAP -- is a treatment in which a mask is worn over the nose and/or mouth while you sleep. The mask is hooked up to a machine that delivers a continuous flow of air into the nose. This air flow helps keep the airways open so that breathing is regular.}
\end{itemize}

\noindent The query requests a definition for ``mechanical ventilation'', but the passage provides details regarding ``Continuous Positive Airway Pressure''.
In our opinion, the passage fails to define ``mechanical ventilation'' and thus does not justify the human-assigned label.
We notice many judgments similarly failing the classification for this query.

Both the above-mentioned cases demonstrate ambiguity that could have been the result of either inaccurate assessments or incomplete information regarding the user intent.

\section{Conclusion}

Continuous efforts are underway to automate the generation of relevance assessments, intending to reduce the cost of manual assessments. 
In our research, we have successfully replicated the results from \citet{bing}, thereby validating the capabilities of OpenAI’s latest model, GPT-4o, in performing relevance assessments automatically. 
Through our experiments with the Deep Learning Tracks from TREC 2019--2023, we have demonstrated the prediction quality of an LLM. 
Moreover, we have demonstrated significant correlations between human and LLM-assessed judgments by evaluating submitted runs to the TREC evaluations. 
Finally, we have repackaged this methodology in an open-source tool, referred to as \umb, that we share with the community.

\begin{acks}
This research was supported in part by the Canada First Research Excellence Fund and the Natural Sciences and Engineering Research Council (NSERC) of Canada.
We'd also like to thank Microsoft for providing access to OpenAI LLMs on Azure via the Accelerating Foundation Models Research program.
\end{acks}

\bibliographystyle{ACM-Reference-Format}
\balance
\bibliography{main}

\end{document}